\pdfoutput=1
\documentclass[final,twocolumn]{elsarticle}
\usepackage{graphicx}
\usepackage{amsmath}
\usepackage{amssymb}

\begin{document}

\begin{frontmatter}

\title{Coulomb Drag and Magnetotransport in Graphene Double Layers}

\author{Seyoung Kim}
\author{Emanuel Tutuc}
\address {Microelectronics Research Center, Department of Electrical and Computer Engineering, University of Texas at Austin, Austin, TX 78758}

\begin{abstract}
We review the fabrication and key transport properties of graphene double layers,
consisting of two graphene monolayers placed in close proximity, independently contacted,
and separated by an ultra-thin dielectric. We outline a simple band structure model relating
the layer densities to the applied gate and inter-layer biases, and show that calculations
and experimental results are in excellent agreement both at zero and in high magnetic fields.
Coulomb drag measurements, which probe the electron-electron scattering between
the two layers reveal two distinct regime: (i) diffusive drag at elevated temperatures,
and (ii) mesoscopic fluctuation-dominated drag at low temperatures. We discuss the
Coulomb drag results within the framework of existing theories.
\end{abstract}

\end{frontmatter}

\section{Introduction}
Closely spaced double layer electron systems possess an additional, layer degree of freedom,
which in certain conditions stabilizes ground states with no counterpart in the single layer
case. Notable examples include fractional quantum Hall states (QHS) at even denominator
fillings, such as $\nu=1/2$ \cite{suen92,eisen92} and $\nu=1/4$ \cite{luhman,shabani}, or a
peculiar QHS at total filling factor $\nu=1$ (layer filling factor 1/2) \cite{yang-moon}.
The $\nu=1$ QHS in interacting double layers displays striking transport properties such
as enhanced inter-layer tunneling \cite{spielman} and counterflow superfluidity \cite{kellogg04,tutuc04},
and has been likened to a BCS exciton condensate \cite{macdonald}. Dipolar superfluidity has been
posited to also occur at zero magnetic field \cite{lozovik} in spatially separated, closely
spaced two-dimensional electron and hole systems, thanks to the pairing of carriers in opposite layers.
Although remarkable progress has been made in the realization of high mobility electron-hole bilayers \cite{croxall,morath},
an unambiguous signature of electron-hole pairing remains to be experimentally observed.
The common thread in these phenomena is the inter-layer Coulomb interaction being comparable in strength
to the intra-layer interaction, leading to many-particle ground states involving the carriers of {\it both} layers.

The emergence of graphene \cite{novoselov2004,novoselov2005,kim2005} as an
electronic material has opened fascinating avenues in the study of the electron
physics in reduced dimensions. Thanks to its atomically thin vertical dimension,
graphene allows separate two-dimensional electron systems to be brought in
close proximity, at separations otherwise not accessible in other heterostructures,
and tantalizing theoretical predictions are based on this property \cite{min,joglekar}.
In light of these observations, it is of interest to explore electron physics in
closely spaced graphene double layers. Here we discuss the fabrication,
and key electron transport properties in this system, namely individual layer resistivity and Coulomb drag.
We introduce a model to describe the layer density dependence on gate and inter-layer bias,
and show that calculations agree well with experimental results in zero and high magnetic fields.
Coulomb drag measurements reveal two distinct regimes: (i) diffusive drag at elevated temperatures,
and (ii) mesoscopic fluctuations-dominated drag at low temperatures.
While we focus here on graphene double layers separated by a thin metal-oxide dielectric,
a system with which the authors are most familiar with \cite{kim2011,kim2012}, we also note recent
progress in graphene double layers separated by hexagonal boron nitride
\cite{ponomarenko,britnell}.

\section{Realization of graphene double layers}

The fabrication of independently contacted graphene double layers starts with the mechanical
exfoliation from natural graphite of the bottom graphene layer onto a 280 nm thick SiO$_2$ dielectric,
thermally grown on a highly doped Si substrate. Electron beam (e-beam) lithography,
metal (Ni or Cr-Au) contact deposition followed by lift-off, and O$_2$ plasma etching are
used to define a Hall bar device. The Al$_2$O$_3$ inter-layer dielectric
is then deposited by atomic layer deposition (ALD), and using an 2 nm thick evaporated
Al film to nucleate the ALD growth. The total inter-layer dielectric thickness for the
samples used our study ranges from 4 nm to 9 nm.  To fabricate the graphene top layer,
a second monolayer graphene is mechanically exfoliated on a SiO$_2$/Si substrate.
After spin-coating poly(metyl metacrylate) (PMMA) on the top layer and curing,
the underlying SiO$_2$ substrate is etched with NaOH, and the top layer along with the
alignment markers is detached with the PMMA membrane. The PMMA membrane is then aligned
with the bottom layer device, and a Hall bar is subsequently defined on the top layer,
completing the graphene double layer.

We focus here on data collected from two samples, labeled 1 and 2, both with a $d=7.5$ nm thick Al$_2$O$_3$
inter-layer dielectric, and with an inter-layer resistance larger than 1 G$\Omega$. The layer mobilities
are $\approx$10,000 cm$^2$/V$\cdot$s for both samples. The layer resistivtities are measured using
small signal, low frequency lock-in techniques as function of back-gate bias (V$_{BG}$), and inter-layer
bias (V$_{TL}$) applied on the top layer. The bottom layer is maintained at the ground (0 V) potential
during measurements. The data discussed here are collected using a pumped $^3$He refrigerator with
a base temperature $T=0.4$ K.

\begin{figure}
\centering
\includegraphics[scale=0.45]{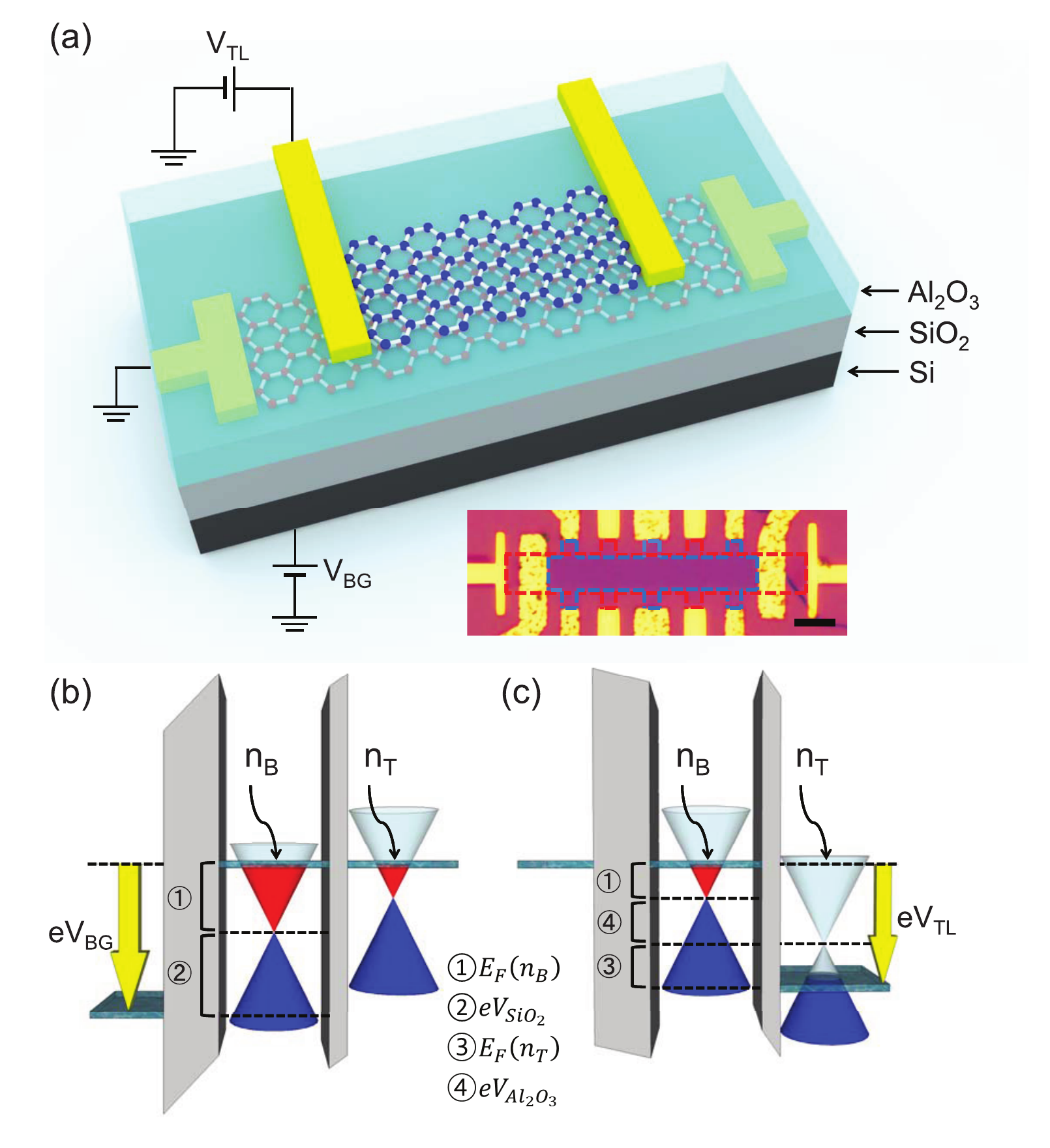}
\caption{(a) Schematic representation of a graphene double layer, consisting of two graphene
monolayers separated by a thin Al$_2$O$_3$ dielectric. The two layer densities can be independently
controlled using the back-gate ($V_{BG}$) bias along with the inter-layer ($V_{TL}$) bias
applied on the top layer. Lower right: optical micrograph of a graphene double-layer device. The
red (blue) contour marks the bottom (top) layer. The scale bar is 5 $\mu$m. (b),(c) Band diagram
of a graphene double layer under an applied back-gate [panel (b)] or inter-layer [panel (c)] bias.
Adapted from Ref. [19].}
\end{figure}

To understand the layer resistivity dependence on gate and inter-layer bias,
it is instructive to examine a band structure model which relates the applied
$V_{BG}$ and $V_{TL}$ biases to the top ($n_{T}$) and bottom ($n_{B}$) layer
densities [Figs. 1(b,c)]. The applied $V_{BG}$ can be written as the sum of the
electrostatic potential drop across the bottom SiO$_2$ dielectric and the Fermi energy of
the bottom layer:
\begin{equation}
eV_{BG}=e^2(n_B+n_T)/C_{SiO_2}+E_F(n_B)
\end{equation}
$E_F(n)$ represents the Fermi energy of graphene relative to the charge neutrality (Dirac) point
at a carrier density $n$; $n$ and $E_F(n)$ are positive (negative) for electrons (holes).
$C_{SiO_2}$ is the SiO$_2$ dielectric capacitance per unit area. Similarly, an applied
$V_{TL}$ can be written as the sum of the electrostatic potential drop across the Al$_2$O$_3$
dielectric, and the Fermi energy of the two layers:
\begin{equation}
eV_{TL}=E_F(n_B)-[E_F(n_T)+e^2n_T/C_{Al_2O_3}]
\end{equation}
A positive $V_{TL}$ applied on the top layer induces electrons (holes) in the bottom (top) layer,
which explains the negative sign for the right hand side terms in Eq. (2). While the above equations implicitly
assume that the two graphene layers are charge neutral at $V_{BG}=V_{TL}=0$ V, a finite doping of the two
layers can be included in the above model using additive constants to the left hand side in Eqs. (1,2).
Alternatively, $V_{BG}$ and $V_{TL}$ can be referenced with respect to the bias values which render
both layers charge neutral, a convention which we adopt in discussing our experimental results \cite{Diracvoltage}.

\begin{figure}
\centering
\includegraphics[scale=0.4]{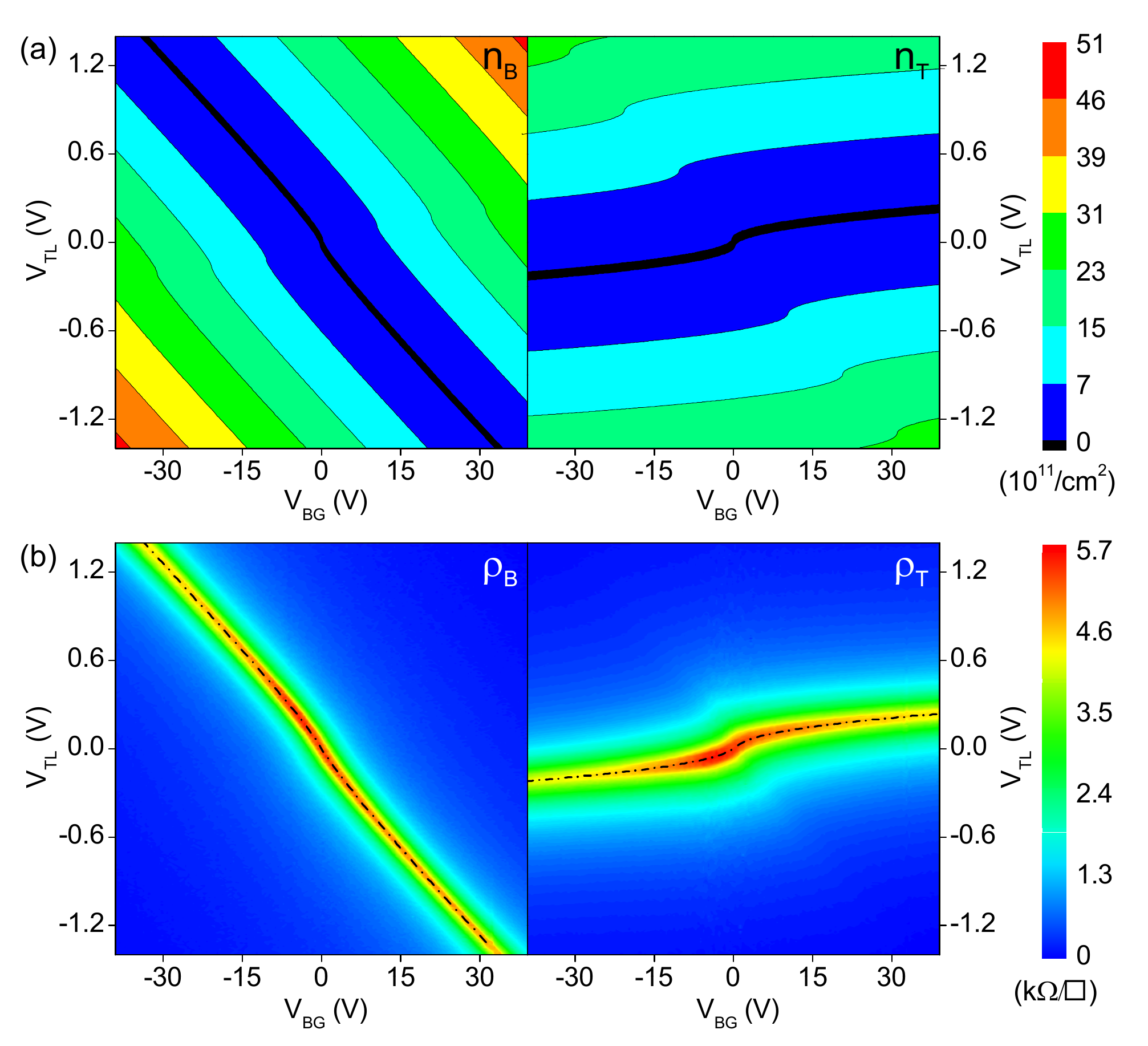}
\caption{(a) Carrier density contour plot of the bottom (left panel) and top (right panel) layer,
calculated as a function of $V_{BG}$ and $V_{TL}$. (b) Longitudinal resistivity measured in the bottom
(left panel) and top (right panel) layer in sample 1 at $T=0.4$ K. Adapted from Ref. [19].}
\end{figure}

\section{Magnetotransport properties of individual layers}

\begin{figure*}
\centering
\includegraphics[scale=0.46]{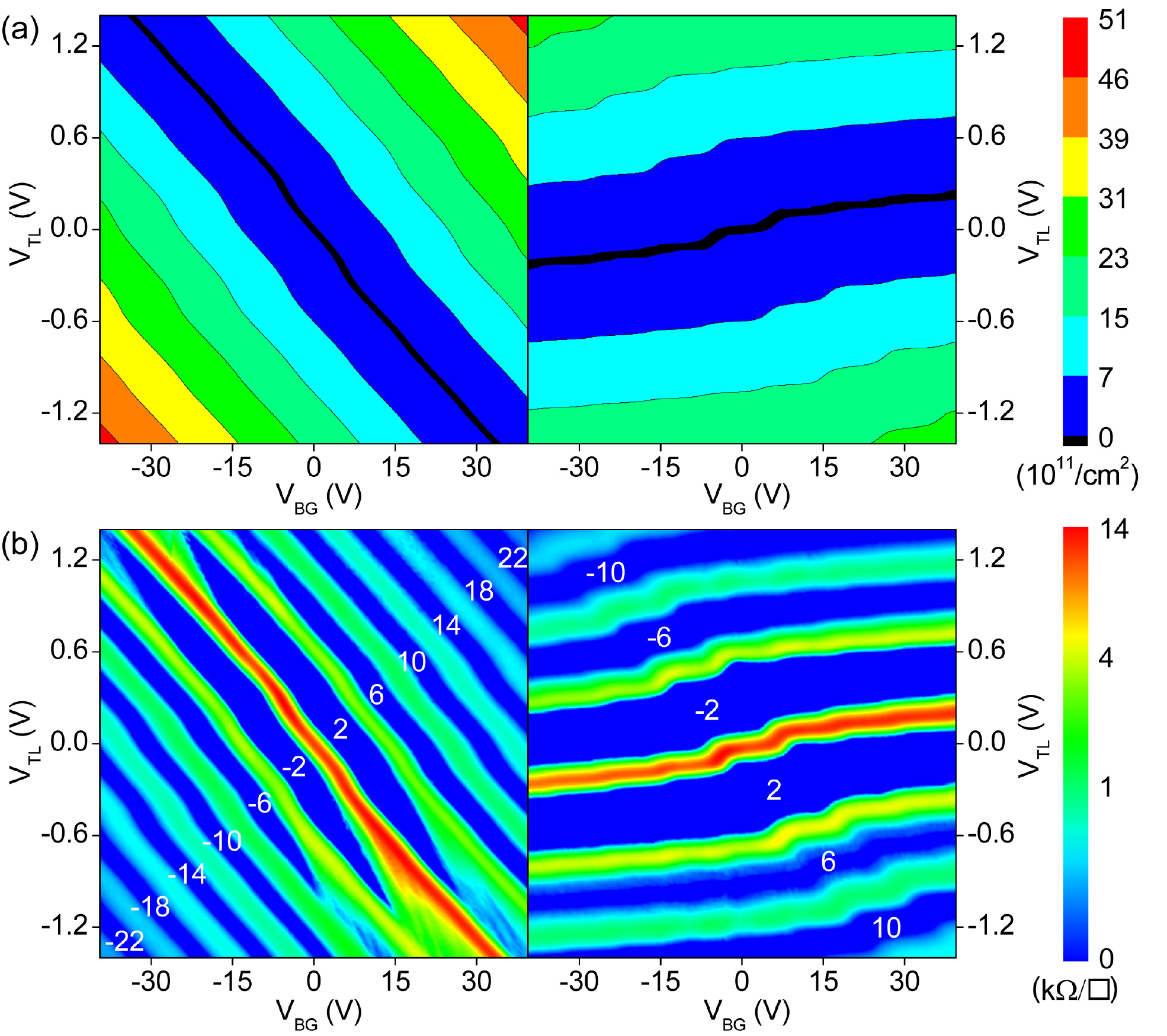}
\caption{(a) Contour plots of $n_B$ (left panel) and $n_T$ (right panel) calculated as a function of $V_{BG}$ and $V_{TL}$ at $B=8$ T.
(b) $\rho_{B}$ (left panel) and $\rho_{T}$ (right panel) measured as a function of $V_{BG}$ and $V_{TL}$ at $B=8$ T and $T=0.4$ K in sample 1.
The data show QHSs marked by vanishing resistivity. The QHS filling factors are marked in both panels.}
\end{figure*}

Figure 2(a) data show contour plots of $n_B$ and $n_T$ calculated according to Eqs. (1,2),
with the following dependence of the Fermi energy on density $E_F(n)=\hbar v_F \sqrt{\pi \cdot n}$, consistent with masseless
particles with a Fermi velocity $v_F$. The bottom SiO$_2$ dielectric capacitance is $C_{SiO_2}=12$ nF$\cdot$cm$^{-2}$,
a value determined experimentally by probing the capacitance of metal pads located in proximity of the graphene double
layer and confirmed by Hall measurements. The inter-layer capacitance value used here is $C_{Al_2O_3}=340$ nF$\cdot$cm$^{-2}$ \cite{capacitance},
and the Fermi velocity $v_F=1.15\times10^8$ cm/s. Figure 2(a) data show that the charge neutrality point of the bottom layer
follows an almost linear dependence on $V_{BG}$ and $V_{TL}$, while the density of the top layer is controlled primarily by the inter-layer bias,
and to a much lesser extent by the back-gate bias. To better understand these observations, let us first neglect $E_F(n)$
in Eqs. (1,2): $n_T$ is controlled by $V_{TL}$ only, and $n_{B}$ depends linearly on $V_{BG}$ and $V_{TL}$,
with $C_{SiO_2}$ and $C_{Al_2O_3}$ respectively as proportionality coefficients. Although this behavior resembles
Fig. 2(a) data, noticeable departures can be observed as a result of the reduced density of states in graphene,
and consequently non-negligible $E_F(n)$. First, $n_T$ does depend on $V_{BG}$, implying an incomplete screening of the back-gate
induced electric field by the bottom layer, or equivalently a finite density of the states in the bottom layer,
hence finite $E_F(n_B)$. Second, the $n_B=0$ line shows a noticeable non-linearity as a function of $V_{BG}$
and $V_{TL}$ near zero gate bias, a consequence of the non-negligible Fermi energy of the top layer.

Figure 2(b) data show contour plots of the longitudinal resistivities $\rho_{B}$ (left panel) and $\rho_{T}$ (right panel)
measured in the bottom and top layers, respectively, at a temperature $T=0.4$ K. A comparison with Fig. 2(a) data reveals a very good agreement
between the experimental data and calculations, validating the model of Eqs. (1,2). Interestingly, this model
can be used to make a direct measurement of the Fermi energy in graphene, by employing one of the two layers as a resistively
detected Kelvin probe \cite{kim2012}. Indeed, setting $n_T=0$ in Eq. (2) yields $eV_{TL}=E_F(n_B)$. The latter relation implies that
the inter-layer bias required to bring the top layer to the charge neutrality point is equal to the Fermi energy of the bottom layer
expressed in units of eV. Consequently, tracking the top layer charge neutrality point in Fig. 2(b) (right panel) yields
the bottom layer Fermi energy as a function of $V_{BG}$.

We next address the layer density dependence on $V_{BG}$ and $V_{TL}$ in a perpendicular magnetic field ($B$).
In the presence of a $B$-field the electrons occupy Landau levels (LL) with energies $E_N=\pm v_F \sqrt{2 \hbar e B |N|}$,
where $N$ is an integer denoting the LL index. Including the spin and valley degrees of freedom each LL has a degeneracy of $4eB/h$.
Quantum Hall states (QHSs) develop at filling factors $\nu=\pm 4(N+1/2)$. Neglecting LL disorder-induced broadening,
the Fermi energy dependence on density therefore writes $E_F=E_N$, where $N=Int[nh/4eB]$ and $Int$ is the nearest integer function.
Figure 3(a) data show contour plots of $n_B$ (left panel) and $n_T$ (right panel) calculated as a function of $V_{BG}$ and $V_{TL}$
at $B=8$ T using Eqs. (1,2), along with the same capacitance values used in Fig. 2(a): $C_{SiO_2}=12$ nF$\cdot$cm$^{-2}$,
$C_{Al_2O_3}=340$ nF$\cdot$cm$^{-2}$. In addition, a finite broadening is assumed for each LL using a Lorentzian density of states distribution
with a width $\gamma_0=14$ meV for the $N=0$ LL and $\gamma_N=6.5$ meV for $|N|>0$ \cite{kim2012}.

Figure 3(b) data show the measured $\rho_{B}$ (left panel) and $\rho_{T}$ (right panel) as a function of $V_{BG}$ and $V_{TL}$ in sample 1,
at $T=0.4$ K. The data display alternating regions of vanishing resistivity corresponding to quantum Hall states, and high resistivity regions
corresponding to half filled LLs. A comparison with Fig. 3(b) data reveals a very good agreement of the charge neutrality point dependence
on $V_{BG}$ and $V_{TL}$ for both layers, effectively validating Eqs. (1,2) model in high magnetic fields. Similar to Fig. 2(b) data,
the $V_{TL}$ value when the top layer is charge neutral is equal to the Fermi energy of the bottom layer. Consequently,
the top layer charge neutrality line of Fig. 2(b) traces the bottom layer Fermi energy as a function of $V_{BG}$,
and shows the staircase dependence corresponding to discrete LLs.

\begin{figure*}
\centering
\includegraphics[scale=0.55]{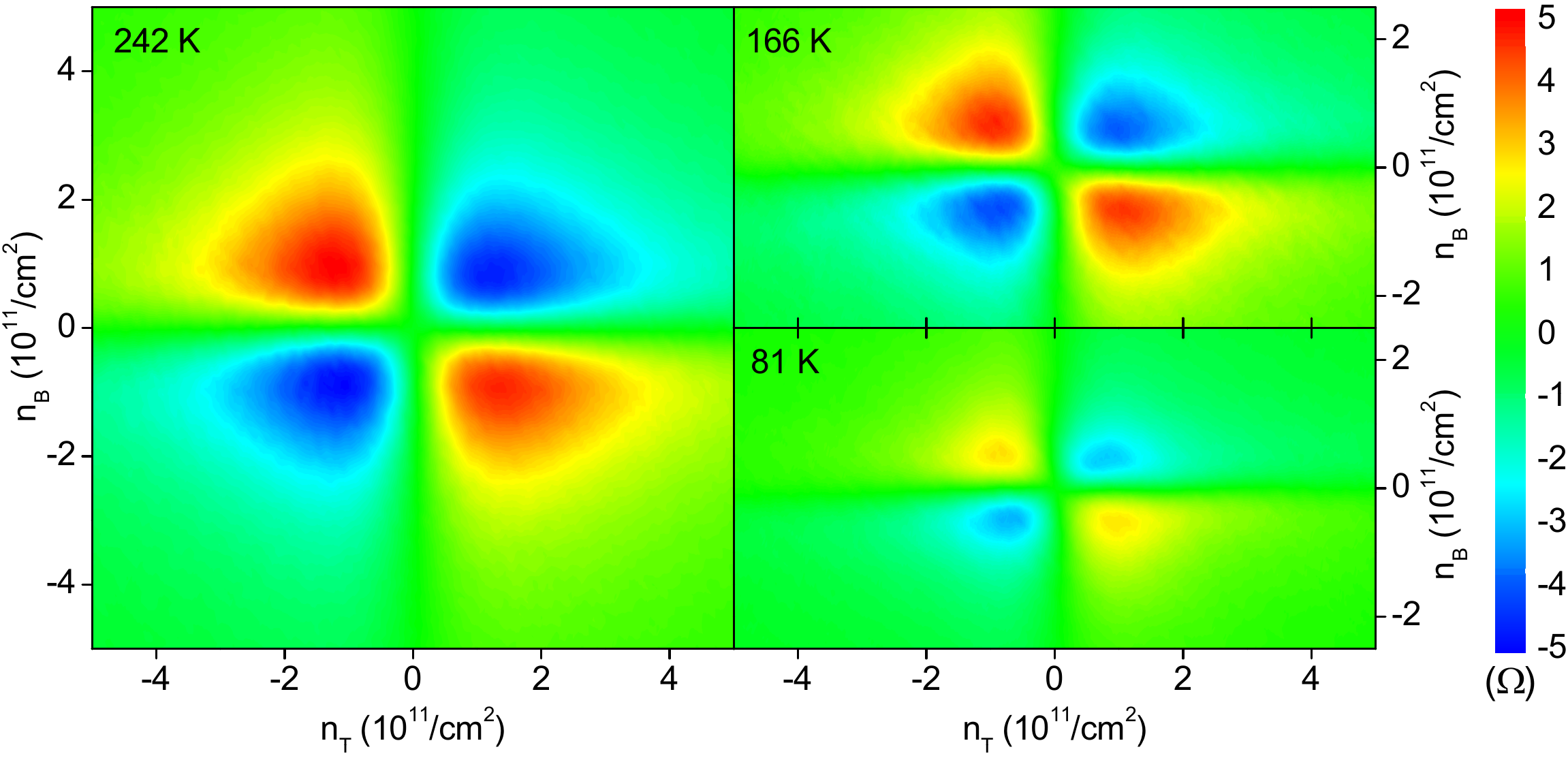}
\caption{Contour plot of $\rho_D$ measured in sample 1 as a function of $n_B$ and $n_T$, and at different $T$-values as indicated
in each panel. The diffusive Coulomb drag shown here increases with $T$, is positive (negative) for opposite (identical)
carrier polarity, and has a similar dependence on layer density in all four quadrants.}
\end{figure*}

\section{Coulomb drag of massless fermions in graphene}

An important attribute of a double layer system is the ability to probe the Coulomb drag between the two layers.
A charge current ($I_{Drive}$) flown in one (drive) layer results in a net momentum transfer to the opposite (drag) layer,
thanks to the Coulomb interaction between electrons in the two layers. If no current is allowed to flow in the drag layer,
a voltage $V_{Drag}$ builds up in order to counter the momentum transfer. The drag resistivity is defined as $\rho_{D}=(V_{Drag}/I_{Drive})\times(W/L)$,
where $L$ and $W$ are the length and width of the sample area over which the drag voltage is measured. In effect, the
drag resistivity is proportional to the scattering rate between electrons located in opposite layers. The dependence
of $\rho_{D}$ on $n_B$, $n_T$, $T$, and the spacing $d$ between the two layers can provide insight into the interaction
between the two layers, as well as the ground state in individual layers.

Coulomb drag between closely spaced carrier systems has been probed in a variety of GaAs/AlGaAs heterostructures which, depending on the type of
sample used, allowed experimental access to electron-electron drag \cite{solomon,gramila}, hole-hole drag \cite{pillarisetty},
and electron-hole drag \cite{sivan,croxall,morath}. Assuming that the ground state in each layer is a Fermi liquid, and that the layer
spacing is sufficiently large such that the Fermi wave-vector $k_F\gg d^{-1}$, the drag resistivity depends on the layer densities,
temperature, and spacing as $\rho_{D} \propto T^2/(n_B^{3/2} \cdot n_T^{3/2} \cdot d^4)$.

\begin{figure*}
\centering
\includegraphics[scale=0.55]{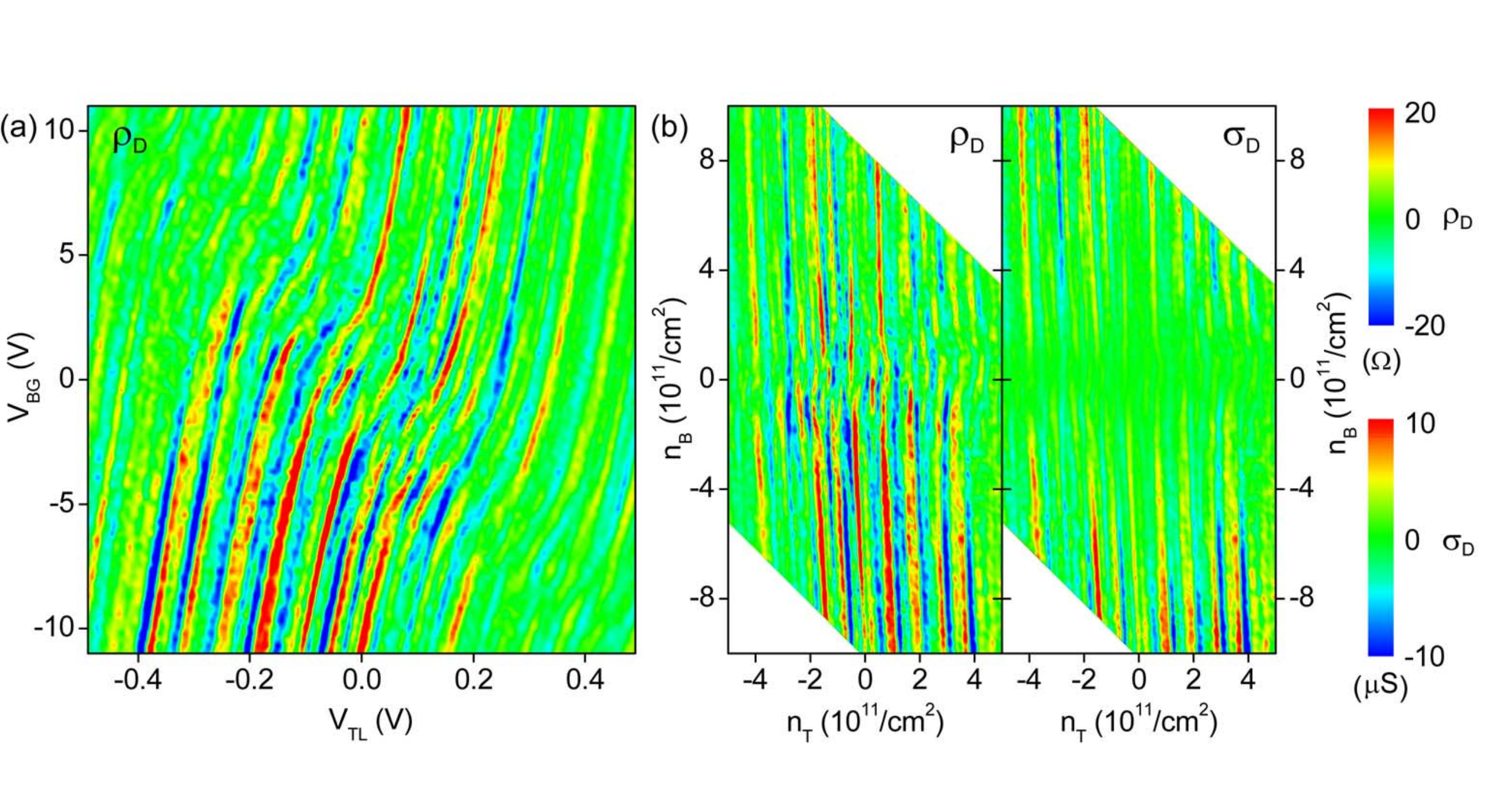}
\caption{(a) Contour plot of $\rho_D$ vs. $(V_{BG},V_{TL})$ measured at $T=4.2$ K in sample 2, using the top (bottom) layer as the drag (drive) layer.
The data shows large mesoscopic fluctuations which follow the constant density lines of the top layer.
(b) Contour plots of $\rho_D$ vs. ($n_{B},n_{T}$) (left panel) and $\sigma_D$ vs. ($n_{B},n_{T}$) (right panel) corresponding to panel (a) data.}
\end{figure*}

In Fig. 4 we show contour plots of $\rho_{D}$ as a function of $n_B$, and $n_T$, measured in sample 1 at different temperatures.
The data were measured by flowing current in the bottom layer, and measuring the drag voltage in the top layer as a function
of $V_{BG}$ and $V_{TL}$, which were subsequently converted into layer densities using Eqs. (1,2). We note that because the drag voltage
is orders of magnitude smaller than the longitudinal voltage drop in the drive layer, very small inter-layer leakage currents
can introduce artifacts in the experimental data; care was taken here to ensure the measured drag voltage is not affected by
inter-layer leakage. Figure 4 data explore $\rho_{D}$ in the four quadrants of the $(n_B,n_T)$ plane, and reveals several noteworthy findings.
First, $\rho_{D}$ is negative (positive) for same (opposite) carrier polarity. Second, $\rho_{D}$ increases with increasing $T$.
Third, $\rho_{D}$ vanishes at large $n_{B,T}$ values, and in the vicinity of the $n_{B,T}=0$ points. Lastly, the $\rho_{D}$ vs. $(n_B,n_T)$ data
in different quadrants are very similar, except for the sign change depending on the carrier type. As we explain below all these observations
are consistent with Coulomb drag of masseless fermions where the ground state in each layer is a Fermi liquid.

Several theoretical studies examined the Coulomb drag in graphene to date \cite{tse,narozhny,peres,hwang,katsnelson,narozhny2011,polini}.
In the weak-coupling limit, defined as $k_F\cdot d\gg 1$, the drag resistance scales as \cite{tse,katsnelson,narozhny2011,polini}
\begin{equation}
\rho_{D}\propto - \frac{h}{e^2} \cdot \frac{T^2}{n_B^{3/2}n_T^{3/2}d^4}
\end{equation}
In the strong-coupling limit, defined as $k_F\cdot d \ll 1$, $\rho_{D}$ is expected to have a weaker dependence on layer density,
and to be independent of the layer spacing as \cite{narozhny2011,polini}
\begin{equation}
\rho_{D}\propto - \frac{h}{e^2} \cdot \frac{T^2}{n_B^{1/2}n_T^{1/2}}
\end{equation}
A close examination of Fig. 4 data reveals a $T$-dependence which is stronger than $\rho_{D}\propto T$,
but weaker than $\rho_{D}\propto T^2$, in the $T$-range probed here. A best fit to $\rho_{D}$ vs. $n_{B,T}$
along the diagonals $n_B=n_T=n$, and $n_B=-n_T=n$ yields $\rho_{D}\propto n^{-\alpha}$, with an exponent
$\alpha=1.7$ little dependent on temperature. The $\alpha$-value is lower than $3$, expected in the weak-coupling
limit, but larger than 1 expected in the strong-coupling limit. This can be readily understood since $k_F \cdot d$
varies between 0.4 and 1 in the density range where $\rho_{D}$ is not vanishingly small. At low densities,
where $k_F \cdot d \ll 1$ disorder-induced puddles in the two layers lead to a decrease of $\rho_{D}$ towards zero.
The experimental $\rho_{D}$ dependence on $n_B$, $n_T$, and $T$ is in good agreement with the expected theoretical
dependence of diffusive Coulomb drag in the Fermi liquid regime, although a quantitative matching between experiment
and theory remains to be established.

As the temperature is reduced below 50 K, the drag resistivity reveals a dramatic departure from the diffusive Coulomb drag of Fig. 4.
In Fig. 5(a) we show a contour plot of $\rho_{D}$ measured as a function of $V_{BG}$ and $V_{TL}$, at $T=4.2$ K in sample 2, and using
the top (bottom) layer as the drag (drive) layer. Unlike the diffusive Coulomb drag of Fig. 4, the data of Fig. 5(a) show a pattern
of large $\rho_{D}$ fluctuations centered around a zero average. The mesoscopic fluctuations increase in amplitude with reducing $T$,
and completely obscure the diffusive drag at low temperatures \cite{kim2011}. The $\rho_{D}$ mesoscopic fluctuations emerge as a result
of phase coherent transport at low temperature, and represent the counterpart of universal conductance fluctuations in Coulomb drag \cite{narozhny2000}.
Similar mesoscopic fluctuations have been observed in Coulomb drag probed between two-dimensional electron systems
in GaAs/AlGaAs heterostructures, albeit at lower temperatures \cite{price2007}.

A comparison of Fig. 5(a) data with the carrier density contour plot of Fig. 2(a) calculations reveal an interesting observation.
The locus of constant $\rho_{D}$ in Fig. 5(a) corresponds to the lines of constant $n_{T}$ in the $(V_{BG},V_{TL})$ plane of Fig. 2(a).
Alternatively stated, the $\rho_{D}$ mesoscopic fluctuations track the drag layer constant density lines in the $(V_{BG},V_{TL})$ plane.
To better illustrate this observation in Fig. 5(b) (left panel) we show a contour plot of $\rho_{D}$ vs. $(n_{B},n_{T})$;
these data represent Fig. 5(a) $\rho_{D}$ data with $(V_{BG},V_{TL})$ axes converted into $(n_{B},n_{T})$ using Eqs. (1,2).
The right panel of Fig. 5(b) (right panel) shows the drag conductivity $\sigma_{D}=\rho_{D}/(\rho_B\cdot\rho_{T})$
vs. $(n_{B},n_{T})$. We note that the $\sigma_D$ fluctuation amplitude has the same order of magnitude as $e^2/h$.
Figure 5(b) data manifestly show that the Coulomb drag mesoscopic fluctuations depend mainly
on the drag layer density ($n_T$), and are largely insensitive to the drive layer density ($n_B$).
This observation is manifestly at variance with the Onsager reciprocity relation, according to which
interchanging the drag and drive layers should not affect the measured $\rho_{D}$.

Examination of Fig. 5(b) data reveals a regular, almost periodic $\rho_{D}$ vs. $n_T$ pattern. To better illustrate
this pattern, in Fig. 6 we show $\rho_{D}$ vs. $n_T$ measured at a fixed drive layer density, $n_B=-5\times10^{11}$ cm$^{-2}$,
and at $T=4.2$ K. The data reveals an almost periodic dependence of $\rho_D$ vs. $n_T$ reminiscent of Fabry-Perot interference \cite{young}.
A simple analysis can relate the top layer Fermi energy ($\Delta E_{F,T}$) change corresponding to $\rho_{D}$ maxima
to a cavity length ($L$), via $\Delta E_{F,T}=h\cdot v_F/L$. A typical value for $\Delta E_{F,T}$ of 10 meV deduced from Fig. 6 data,
corresponds to $L=0.5$ $\mu$m, a value comparable to the phase coherence length in graphene on SiO$_2$ \cite{berezovsky} at $T=4.2$ K.

\begin{figure}
\centering
\includegraphics[scale=0.18]{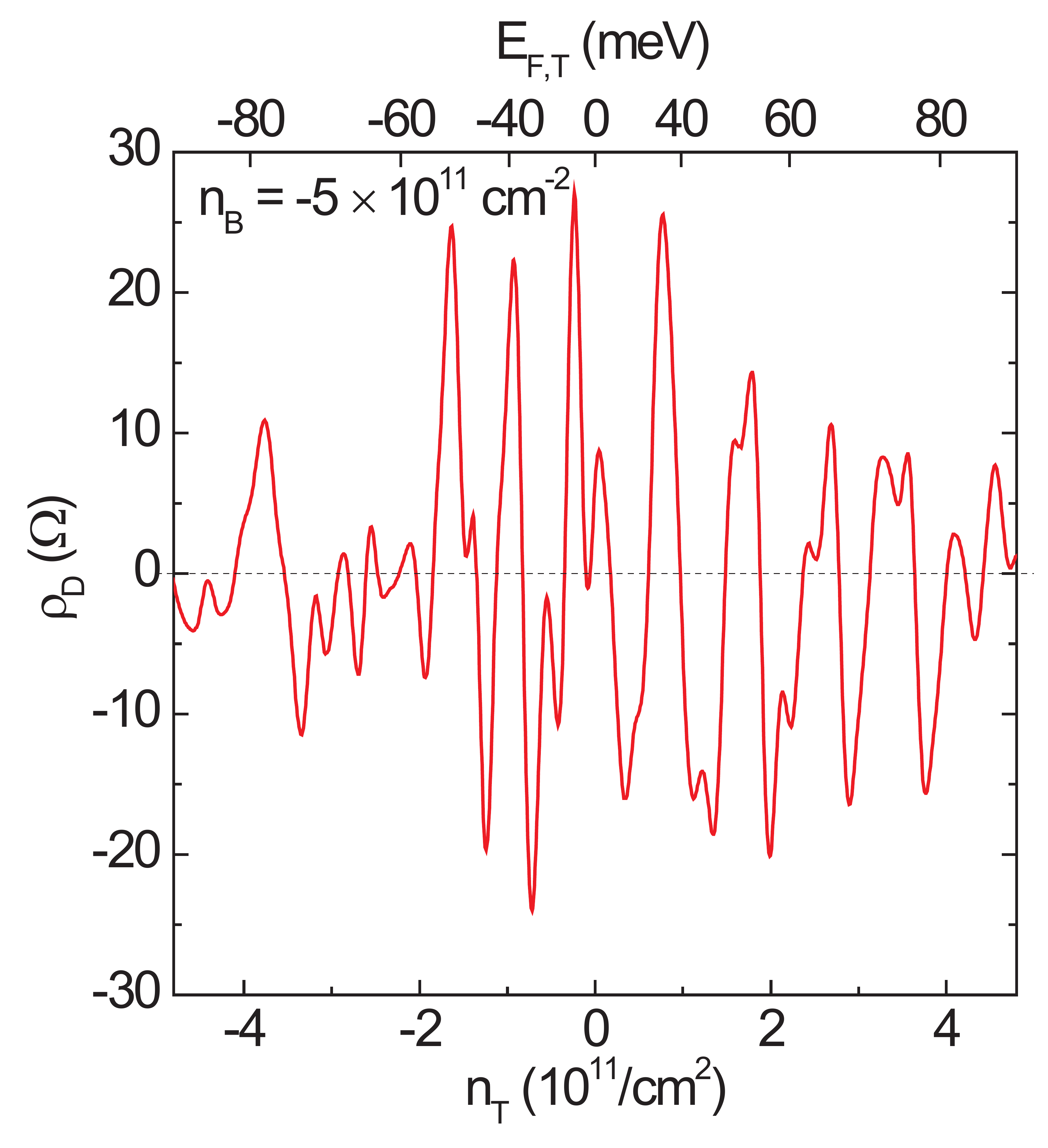}
\caption{$\rho_D$ vs. $n_T$ measured at a fixed bottom layer density $n_B=-5\times10^{11}$ cm$^{-2}$ in sample 2,
at $T=4.2$ K and using the top (bottom) layer as the drag (drive) layer. The top axis represents the Fermi energy
of the drag layer. The data show an almost periodic fluctuation pattern.}
\end{figure}

\section{Conclusions}

In summary, we present a magnetotransport and Coulomb drag study in graphene double layers, consisting of two independently
contacted graphene monolayers separated by a thin Al$_2$O$_3$ dielectric. The Coulomb drag probed in this system reveals two regimes:
(i) diffusive drag at elevated ($T>50$ K) temperatures, and (ii) mesoscopic fluctuations dominated drag at low temperatures.
The temperature dependence of the diffusive drag is consistent with the Fermi liquid theory, while the density dependence
suggests the layers are close to the strong-coupling regime \cite{narozhny2011,polini}. The Coulomb drag mesoscopic fluctuations
observed at low temperature depend mainly on the drag layer density, and are largely insensitive to the drive layer density,
an observation which is at variance with the Onsager reciprocity relation.

We thank A. H. MacDonald, B. Narozhny, N. M. R. Peres, W.-K. Tse, S. K. Banerjee, and L. F. Register for
discussions. We gratefully acknowledge support from ONR, NRI, and NSF (DMR-0819860).


\begin{thebibliography}
\bibitem{Diracvoltage}
The the bias values at which both layers are at the Dirac point are
$V_{BG}=8$ V, $V_{TL}=0.01$ V for sample 1, and $V_{BG}=6.8$ V, $V_{TL}=-0.1$ V
for sample 2.
\end{thebibliography}
\end{document}